\documentclass[bib]{statapress}

\usepackage[crop,newcenter,frame]{pagedims}

\usepackage{sj}

\usepackage{epsfig}

\usepackage{stata}

\usepackage{amsmath}

\usepackage{shadow}

\usepackage[nameinlink]{cleveref}

\sjsetissue{$vv$}{$ii$}{$mm$}{$yyyy$}

\usepackage{threeparttable} 
\usepackage{tabularx}

\begin{document}

\inserttype[st0001]{article}
\author{David Roodman}{%
	David Roodman\\Open Philanthropy\\San Francisco, CA\\david.roodman@openphilanthropy.org
}
\title[Julia as a universal platform for statistical software development]{Julia as a universal platform for statistical software development}
\maketitle

\begin{abstract}
The {\tt julia} package integrates the Julia programming language into Stata. Users can transfer data between Stata and Julia, issue Julia commands to analyze and plot, and pass results back to Stata. Julia's econometric ecosystem is not as mature as Stata's or R's or Python's. But Julia is an excellent environment for developing high-performance numerical applications, which can then be called from many platforms. For example, the {\tt boottest} program for wild bootstrap--based inference \citep{RMNW_2019} and {\tt fwildclusterboot} for R \citep{fwildclusterboot} can use the same Julia back end. And the program {\tt reghdfejl} mimics {\tt reghdfe} \citep{Correia2017:HDFE} in fitting linear models with high-dimensional fixed effects while calling a Julia package for tenfold acceleration on hard problems. {\tt reghdfejl} also supports nonlinear fixed-effect models that cannot otherwise be fit in Stata---though preliminarily, as the Julia package for that purpose is immature.

\end{abstract}

\newpage

\section{Introduction}
\label{intro}

Whenever you issue a command to perform a statistical calculation, you initiate a chain of events of unfathomable complexity: your request cascades through layers of code crafted over decades by thousands of programmers in multiple languages, as it is expanded into millions or billions of computational steps---steps that are executed, if you're lucky, in the time it takes to read this sentence.

In Stata, for example, a call to {\tt regress} or {\tt gsem} starts out in the ado programming language. The ado code translates your command into a digital problem definition and passes that to routines written in C/C++ or Stata's built-in Mata language. In modern Stata, an ado program can even call Python or Java. The called code will probably invoke the Fortran-based LAPACK library to invert matrices or solve linear systems.

Here I introduce a package that forges a fresh link between Stata and another language, Julia. Julia is new: it reached version 1.0 in 2018. It is also free, and available for Windows, Linux, and macOS. Like Python code, Julia code is fully portable across those platforms and can run as soon as downloaded.

For creators of computationally intensive software, Julia has many strengths. More so than with Stata (but less so than with Python), if you need a tool to perform a fundamental task such as exploiting GPUs, there's a good chance someone has written and posted it. Julia's creators set out to solve the ``two-language problem," the inefficiency that arises when programmers have to split a project between high- and low-level languages as they trade off ease of software development for speed of execution. Python and C++ is a common two-language combination, as is R and C++. Julia resembles Python in syntax, but was built from the start for just-in-time compilation, meaning that it translates Julia functions to machine language when they are first called, and stores that machine code for reuse. A final strength is that, like Lisp, Julia can treat code as data. This feature opens the door to meta-programs that can automatically optimize one's code by rewriting it before compilation, or generate new code to take derivatives of complex calculations already implemented.

Julia has disadvantages. While the Julia ecosystem contains the foundation for statistical work---packages to manage data sets, fit common models, and generate plots---many packages are immature or poorly documented. Standards for storing and returning results are less well developed than in Stata. Much less statistical functionality is available than in Stata or R, or even Python.

Since Julia is a great home for numerical software, but not (yet) a great home for the end user, a sweet spot for the language is back end development. Methods of estimation and inference can be implemented in Julia, and called from Stata, R, and other platforms. The core can be augmented and optimized in one place, rather than separately on each platform.

The {\tt julia} package links Stata to Julia. It follows in the footsteps of projects to link statistical software to other languages. StataCorp introduced the Stata plugin interface (SPI) for C/C++ in Stata 8.1, in 2003. It added the built-in Mata language in Stata 9. \citet{Fiedler_2012} envisioned running Python inside of Stata. Following the suggestion of a StataCorp employee, \citet{Fiedler_2013} partly realized that vision by writing a C++ plugin for Stata. Perhaps inspired by Fiedler, Stata 16 fully integrated Python. Java plugins became possible in Stata 13 and the Java prompt debuted in the Stata results window in version 17. \citet{Haghish2019} introduced {\tt rcall}, which allows one to issue R commands from Stata; it passes content between the two systems by saving files to disk. Outside of Stata, the JuliaCall \citep*{JuliaCall} and JuliaConnectoR \citep{JuliaConnectoR} packages link R to Julia. Another package named JuliaCall does the same for Python.\footnote{See \url{juliapy.github.io/PythonCall.jl/stable}.}

The {\tt julia} package was most inspired by \citet{Fiedler_2013}. Its core too is a C++ plugin, which comes precompiled for Windows, Linux, and macOS with Intel or Apple silicon. From the user's point of view, the package has three components:

\begin{itemize}
	\item The {\tt jl} command for Stata, which prefixes single Julia lines or, when typed by itself, starts a Julia session in Stata.
	
	\item {\tt jl} subcommands for managing Julia packages and transferring data between Stata and Julia.
	
	\item Julia functions to read and write Stata macros, scalars, matrices, and variables.
\end{itemize}

The package is designed to hide the complexities of the Stata-Julia linkage, so that getting started with Julia requires nothing more than installing the {\tt julia} package for Stata.

Two user-written Stata packages call Julia via {\tt julia}. The {\tt boottest} program for wild bootstrap--based inference \citep{RMNW_2019} now accepts a {\tt julia} option, which causes it to use my Julia package WildBootTests.jl rather than the Mata back end I originally wrote for {\tt boottest}. And the new {\tt reghdfejl} mimics the {\tt reghdfe} and {\tt ivreghdfe} programs for fitting linear models with high-dimensional fixed effects (HDFE) \citep*{Correia2017:HDFE}. {\tt reghdfejl} calls the Julia package FixedEffectModels.jl, whose lead developer is Matthieu Gomez. {\tt reghdfejl} also has the ability, in beta, to fit nonlinear HDFE models, such as the logit and negative binomial, via Johannes Boehm's GLFixedEffectModels.jl. In the last case, using Julia brings new functionality to Stata. In all cases, calling Julia cuts runtime on hard problems, sometimes by an order of magnitude.

Stata users can thus benefit from Julia and {\tt julia} without using them directly. Still, this paper explains the new {\tt julia} package. It starts by demonstrating how to run Julia from Stata, in \cref{firstsession}. \Cref{applications} introduces applications: it introduces {\tt reghdfejl} and the Julia back end for {\tt boottest}; and it demonstrates graphing via Julia. \Cref{tips} gives tips and warns of some limitations. \Cref{syntax} more thoroughly describes the package's Stata commands and Julia functions.

\section{Hands-on introduction}
\label{firstsession}

\subsection{Installation and first examples}
\label{sec:install}

To install the {\tt julia} package in Stata, type:

\begin{stlog}
	ssc install julia
\end{stlog}
You can now call Julia from Stata:

\begin{stlog}
	. jl: "Hello world!"
	\oom
	Starting Julia
	"Hello world!"
\end{stlog}
This operation may take several minutes. In Windows, mysterious console windows may appear and disappear. Julia and the Julia version manager will be installed, if they haven't been already. The version manager will be configured to guarantee that Julia 1.11, the version currently required by {\tt jl}, is always available. {\tt jl} will also install two Julia packages, DataFrames.jl and CategoricalArrays.jl, along with all the packages {\it they} depend on. DataFrames.jl \citep{JSSv107i04} is the dominant Julia framework for data sets, meaning rectangular grids whose columns have names and potentially different data types. CategoricalArrays.jl implements the Julia equivalent of Stata variables with data labels---ones coded, for example, so that 0 means ``black" and 1 means ``white." (Both packages are hosted on GitHub and are maintained by users, who accept improvements from others.)

While Julia's developers work hard to assure that programs written in older versions work in newer ones too, they do not guarantee backward compatibility in Julia's low-level, C-based interface, which is what the {\tt julia} package for Stata links to. Changes to that interface can cause {\tt jl} to completely crash Stata. (Ask me how I know.) This is why {\tt jl} is designed to work \textit{only} with the latest version of Julia at this writing, 1.11. Fortunately, Julia's version manager makes it easy to maintain several Julia versions at once on a computer. So if you want to use a different version outside of Stata, {\tt jl}'s need for 1.11 won't interfere.\footnote{If revisions to Julia's innards break the {\tt julia} package for Stata, I expect to revise the package to stay up-to-date. The required Julia version may change.}

In general when using Julia, you will experience delays at four points: when packages are installed; when they are first used; when they are first used in a Julia session; and when particular features within them are first used, such as clustering standard errors. A thrust of recent development in the Julia language has been to move the behind-the-scenes processing to earlier stages in order to shorten the lags during regular use.

In the session below, I load Stata's ``auto" data set, extract five variables, run a regression on them in Stata, make sure the package I will use for estimation in Julia is installed on my computer, and then copy the data to a Julia DataFrame named {\tt auto}:

\begin{stlog}
. sysuse auto
(1978 automobile data)
{\smallskip}
. keep price headroom foreign mpg turn
{\smallskip}
. reg price headroom foreign\#c.mpg, cluster(turn)
{\smallskip}
Linear regression                               Number of obs     =         74
                                                F(3, 17)          =       9.69
                                                Prob > F          =     0.0019
                                                R-squared         =     0.2939
                                                Root MSE          =       2531
{\smallskip}
                                   (Std. err. adjusted for 18 clusters in turn)
\HLI{14}{\TOPT}\HLI{64}
              {\VBAR}               Robust
        price {\VBAR} Coefficient  std. err.      t    P>|t|     [95\% conf. interval]
\HLI{14}{\PLUS}\HLI{64}
     headroom {\VBAR}  -273.2403   274.4352    -1.00   0.333     -852.248    305.7674
              {\VBAR}
foreign#c.mpg {\VBAR}
    Domestic  {\VBAR}  -344.0717   72.40801    -4.75   0.000    -496.8392   -191.3041
     Foreign  {\VBAR}  -267.0448   57.09596    -4.68   0.000    -387.5067   -146.5828
              {\VBAR}
        _cons {\VBAR}   13743.64     2176.8     6.31   0.000      9150.99    18336.28
\HLI{14}{\BOTT}\HLI{64}
{\smallskip}
. jl AddPkg FixedEffectModels
\smallskip
. jl save auto
Data saved to DataFrame auto in Julia
\end{stlog}
{\tt jl AddPkg} and {\tt jl save} are not Julia commands. They are subcommands of the Stata command {\tt jl}, which is part of the {\tt julia} package. If an object named {\tt auto} already exists in the Julia environment, it is overwritten by the incoming copy of the Stata variables.

Next I type {\tt jl} by itself, which changes the prompt the same way that typing {\tt mata} or {\tt python} or {\tt java} does. It starts an interactive Julia session, in which I can type multiple Julia commands. I start with displaying the new Julia DataFrame by typing its name:

\begin{stlog}
. jl
\HLI{24} Julia (type {\bftt{exit()}} to exit) \HLI{24}
jl> . auto
74×5 DataFrame
 Row {\VBAR} price   mpg     headroom  turn    foreign
     {\VBAR} Int16?  Int16?  Float32?  Int16?  Cat…?
\HLI{5}{\PLUS}\HLI{73}
   1 {\VBAR}   4099      22       2.5      40  Domestic
   2 {\VBAR}   4749      17       3.0      40  Domestic
   3 {\VBAR}   3799      22       3.0      35  Domestic
   4 {\VBAR}   4816      20       4.5      40  Domestic
   5 {\VBAR}   7827      15       4.0      43  Domestic
   6 {\VBAR}   5788      18       4.0      43  Domestic
   7 {\VBAR}   4453      26       3.0      34  Domestic
   8 {\VBAR}   5189      20       2.0      42  Domestic
     {\VBAR}                                      
  68 {\VBAR}   3748      31       3.0      35  Foreign
  69 {\VBAR}   5719      18       2.0      36  Foreign
  70 {\VBAR}   7140      23       2.5      36  Foreign
  71 {\VBAR}   5397      41       3.0      35  Foreign
  72 {\VBAR}   4697      25       3.0      35  Foreign
  73 {\VBAR}   6850      25       2.0      36  Foreign
  74 {\VBAR}  11995      17       2.5      37  Foreign
                                       59 rows omitted
\end{stlog}
Along the top are the column names, and below them the types. {\tt Int16} means a 16-bit integer, equivalent to Stata {\tt int}. {\tt Float32} corresponds to Stata {\tt float}. The question marks at the ends of the type names indicate that the types have been modified to allow the special value {\tt missing}. By default, {\tt jl save} makes all the columns accept missing.

{\tt jl save} also converts the {\tt foreign} variable, which is coded as 0/1 in Stata and has value labels ``Domestic" and ``Foreign," to a {\tt CategoricalVector}. The type name ``CategoricalVector?" does not fit in the column head, so it is shortened to ``Cat...?". While the Julia version of {\tt foreign} is also represented internally as a column of integers, it \textit{presents} as a text variable. All commands involving the Julia variable would refer to display values such as ``Domestic," not to numeric codes as in Stata.

Next I run the same regression as before, but now in Julia, with the {\tt reg()} function of the Julia package FixedEffectModels.jl:

\begin{stlog}
jl> . using FixedEffectModels      # load package for use
\smallskip
jl> . reg(auto, term(:price) ~ term(:headroom) + term(:foreign){\&}term(:mpg),
  ...     Vcov.cluster(:turn))
FixedEffectModel                               
=========================================================================================
Number of obs:                           74   Converged:                             true
dof (model):                              3   dof (residuals):                         16
R²:                                   0.294   R² adjusted:                          0.264
F-statistic:                         9.6942   P-value:                              0.001
=========================================================================================
                          Estimate  Std. Error     t-stat  Pr(>|t|)  Lower 95
\HLI{89}
headroom                  -273.24      274.435  -0.995646    0.3342   -855.017    308.536
foreign: Domestic & mpg   -344.072      72.408  -4.75185     0.0002   -497.57    -190.574
foreign: Foreign & mpg    -267.045      57.096  -4.67712     0.0003   -388.083   -146.007
(Intercept)              13743.6      2176.8     6.31369     <1e-04   9129.03   18358.2
=========================================================================================
\end{stlog}
The Stata and Julia results match. But the Julia regression command needs explaining. First some cosmetic comments:
\begin{itemize}
\item Because the regression command is long, I break it in two. After I hit the enter key at the end of the first line, {\tt jl} detects that the line is not grammatical by itself and prints ``{\tt ...}" to prompt me to continue.

\item Stata has the {\tt \#} and {\tt \#\#} interaction operators. The corresponding symbols in Julia are {\tt \&} and {\tt *}. The first appears above.

\item All variable names in the command line are prefixed with ``{\tt :}". That indicates that the names are {\it symbols}, not stand-alone Julia objects like the DataFrame {\tt auto}. The symbols are names of columns in {\tt auto}.

\item Because {\tt foreign} is stored as a {\tt CategoricalVector}, {\tt reg()} automatically treats it as a factor variable, generating a pair of dummies to be interacted with {\tt mpg}. There is no need for a Julia equivalent of Stata's ``{\tt i.}'' prefix. However, in Julia, treating non-{\tt CategoricalVectors} as factor variables requires a more awkward syntax. One must pass {\tt reg()} an argument with a dictionary mapping the variables to coding types, such as with {\tt contrasts=Dict(:headroom=>DummyCoding())}.
\end{itemize}

The command line contains deeper complexity. In Julia, as in R and Python, every regression command is a statement in a full-fledged programming language. This tends to complicate the syntax for even basic operations. It is hard to implement an R or Python line as clean as {\tt regress y x}. In the Julia command above, {\tt term(:price)}, {\tt term(:headroom)}, ..., etc., create instances of the object type {\tt Term} to represent variables within a statistical model. The {\tt Term} structure is defined in the StatsModels.jl package, which FixedEffectModels.jl loads and makes available to the user. The content of each {\tt Term} is just a symbol, such as {\tt :price}. What give {\tt Term}'s life are the operations that StatsModels.jl defines \textit{on} them. The {\tt +} operator is repurposed (``overloaded'') to form lists (tuples) of {\tt Term}'s. The {\tt $\sim$} operator binds together two {\tt Term} tuples to make a {\tt FormulaTerm}, a structure whose distinctive feature is having designated left and right sides, just as in a linear regression model. In the example, an entire {\tt FormulaTerm} is passed as the second argument to {\tt reg()}. In this way, the primitives of a flexible programming language are composed to articulate a statistical model.

This standard for representing regression models is not intrinsic to Julia. It has emerged from the work of users. And it is transparent. In Stata, one could type {\tt jl: dump(term(:c) $\sim$ term(:a) + term(:b))} to reveal the contents of a {\tt FormulaTerm}. The software that builds and applies these structures is on GitHub, all written in Julia. The StatsModels.jl formula language is also extensible\footnote{See \url{juliastats.org/StatsModels.jl/stable/internals/\#extending}.}. As we will see, FixedEffectModels.jl extends the formula language in order to mark variables whose fixed effects are to be absorbed.

Yet a powerful feature of Julia makes it practical to ignore most of this complexity. The regression command above can be simplified with a \textit{macro}. Where macros in Stata are strings, in Julia they are code for manipulating code: meta-programs. StatsModels.jl defines a macro called {\tt @formula}, which lets one write models more simply. Thus, I can write:

\begin{stlog}
jl> . m = reg(auto, @formula(price ~ headroom + foreign{\&}mpg), Vcov.cluster(:turn))
\oom
\end{stlog}
This is how regressions are usually specified in Julia. Before this new command is compiled and executed, {\tt @formula} turns it into something closer to the earlier regression command.\footnote{Johannes Boehm's experimental Douglass.jl package includes macros to translate Stata-like commands into Julia ones. An example: {\tt d"bysort :Species : egen :z = sum(:SepalLength) if :SepalWidth .> 3.0"}.}

Notice one other change to the regression command: it now starts with {\tt m =} in order to store the fitted model in the Julia variable {\tt m}. This lets me extract return values and pass them back to Stata:

\begin{stlog}
jl> . st\_numscalar("adjR2", adjr2(m))
jl> . exit()
\HLI{84}
. display "Adjusted R2 = " adjR2
Adjusted R2 = .26365506
\end{stlog}
The Julia function {\tt st\_numscalar()} is part of the {\tt julia} package; like its Mata namesake, it reads and writes Stata scalars.

\subsection{Built for speed}

To partially convey why Julia is a good environment for scientific computing, I present another example. Suppose we have a data set with 10 million rows and 10 columns named {\tt x1-x10}. Given a $10\times10$ matrix ${\bf Q}$, we want to compute the norm of each row according to the quadratic form defined by ${\bf Q}$ and store the results in a new column. That is, for each row $x_i$, now viewed as a column vector, we want $x_i' {\bf Q} x_i$. It is natural to implement this computation with triply nested {\tt for} loops. That's a good way to do it in Julia (inside Stata):

\begin{stlog}
jl AddPkg LoopVectorization  // one-time package installation from GitHub
jl: using LoopVectorization  // load package for use now
\smallskip
jl: function XQX(Q, X)                             ///
      N, M = size(X);                              ///
      retval = zeros(N);                           ///
      @tturbo for i in 1:N                         ///
        for j in 1:M                               ///
          for k in 1:M                             ///
            retval[i] += X[i,j] * Q[j,k] * X[i,k]  ///
          end                                      ///
        end                                        ///
      end;                                         ///
      return retval                                ///
    end
\end{stlog}
Unlike in Python, indentation has no semantic meaning; this code snippet is indented only for readability.

We saw in the previous example that one can start an interactive Julia session inside Stata by typing {\tt jl} (and end it with {\tt exit()}). But the interactive mode is not accessible in do files. To exemplify Julia code in a do file, the above snippet is therefore written effectively as a single line. It is typographically, not syntactically, split across multiple lines with Stata's continuation token, {\tt ///}. When putting many Julia commands on what is syntactically one line, most Julia commands not immediately followed by {\tt end} need to terminate with a semicolon, as illustrated above.

The code for this new Julia function, {\tt XQX()}, is mostly straightforward. It initializes the return value to a column of 0's, performs the calculation through nested loops, and returns the result. One detail is unusual: another macro is invoked. {\tt @tturbo} is part of Chris Elrod's LoopVectorization.jl package. {\tt @tturbo} analyzes the code that follows it and substantially---but invisibly---rewrites it for speed. It may unroll a loop to reduce the number of jumps back to the top. The macro, or the compiler, may spot the opportunity to move {\tt X[i,j] *} out of the inner loop since it does not depend on the loop's index, {\tt k}. The macro may even reorder the nesting of the loops with an eye toward the fact that in Julia, matrices are stored column by column, and on-chip memory caching makes it is faster to access adjacent memory locations in sequence. It may tailor the code so that the resulting machine language exploits the ability of modern chips to vectorize operations {\it within} each CPU core, in 256- or 512-bit registers---what is called single instruction, multiple data (SIMD) execution. And {\tt @tturbo} will exploit multithreading, the subdivision of work \textit{across} cores; that is what the first ``t" in {\tt tturbo} stands for.

Triply nested implementations in Mata and Python are much slower, I assume because the calculations in the innermost loop have to be interpreted a billion times. In constrast, Julia translates all the code down to machine language just once. The best alternatives I have found in the other two languages are essentially the same:

\begin{stlog}
mata
function XQX(Q, X)
  return (rowsum((X * Q) :* X))
end

python
def XQX(Q,X):
  return ((X @ Q) * X).sum(axis=1)
end
\end{stlog}
Both delegate the computation to built-in matrix operations, which are fast because they are implemented in compiled C or Fortran. However, both contain a subtle inefficiency. They create the large, temporary matrices ${\bf XQ}$ and ${\bf XQ \otimes\bf X}$, before summing their rows and dispensing with them. Since in our example ${\bf X}$ is $10^7 \times 10$, and since double-precision numbers take 8 bytes, the temporary matrices require some 800 megabytes each. Allocating and deallocating memory itself takes time. Expanding the memory footprint of an algorithm can also reduce speed by exceeding the capacity of a CPU's memory caches, forcing the CPU to idle while data are transferred to and from the computer's main memory. In contrast, the triply-looped Julia code demands essentially no temporary storage.

This inefficiency of Mata and Python can be seen as an artifact of the two-language problem Julia was designed to solve. The best one can do in those languages is rely on a set of canned routines, such as for matrix multiplication, that are written in another language (C++ or Fortran), routines that, while fast on their own terms, are not optimized for the purpose at hand and nearly impossible to revise.

To create a test bed for these functions, I construct ${\bf X}$ as a Stata data set and ${\bf Q}$ as a Stata matrix, then copy them into Mata, Python, and Julia. The {\tt set rmsg on} command generates timing reports:

\begin{stlog}
. python: import numpy as np
. python: from sfi import Data, Matrix
\smallskip
. set obs 10000000
Number of observations (\_N) was 0, now 10,000,000.
\smallskip
. drawnorm double x1-x10
. mata  : st\_matrix("Q", makesymmetric(runiform(10,10)))
\smallskip
. mata  : Q = st\_matrix("Q")
. python: Q = np.asarray(Matrix.get("Q"))
. jl    : Q = st\_matrix("Q");
\smallskip
. set rmsg on
\smallskip
. mata: X = st\_data(.,"x1-x10")
r; t=0.07
\smallskip
. python: X = np.asarray(Data.get(var="x1 x2 x3 x4 x5 x6 x7 x8 x9 x10"))
r; t=6.73
\smallskip
. jl: X = st\_data("x" .* string.(1:10));
r; t=0.24
\smallskip
. jl: X = st\_data("x" .* string.(1:10));
r; t=0.15
\end{stlog}
Under a Julia convention, the semicolons at the end of the {\tt jl} lines suppress printing of the results of the assignments. The Julia data copying command is run twice in order to show that {\tt st\_data()}, which is also part of the {\tt julia} package, takes extra time on first use because that is when it is compiled down to machine code. 

We see that on a Windows laptop with an Intel i9-13900H processor, importing the 100 million data points into a matrix takes just 0.07 seconds for Mata, 0.15 seconds for Julia (the second time), and a substantial 6.73 seconds for Python. In {\tt jl}, the data copying is performed by C++ code that is multithreaded across columns.

Next I test the three functions I defined:

\begin{stlog}
. mata: y = XQX(Q,X)
r; t=0.31
\smallskip
. python: y = XQX(Q,X)
r; t=0.48
\smallskip
. jl: y = XQX(Q,X);
r; t=1.39
\smallskip
. jl: y = XQX(Q,X);
r; t=0.05
\end{stlog}
The Julia function is especially slow on first use because that is when {\tt @tturbo} performs its magic. On later calls, Julia is fast: it takes 0.05 seconds, which works out to about 40 billion floating point operations per second. Mata takes 0.31 seconds and Python, 0.48. To reduce the burden of the first call, if the Julia {\tt XQX()} function were part of a shared package, it could be accompanied by a script much like the test above. The Julia package manager would automatically run the script after installing the package in order to trigger immediate compilation. Julia would save the generated machine code for future use.

\section{Applications}
\label{applications}

This section presents three applications of {\tt julia}. They involve HDFE modeling, wild bootstrap--based inference, and plotting.

\subsection{Estimation with high-dimensional fixed effects}

The introduction of {\tt reghdfe} in 2014 \citep*{Correia2017:HDFE} was an important event in applied econometrics. Building on \citet{GP2010} and \citet{Gaure2010}, the Mata-based program proved the computational tractability of linear models with many sets of fixed effects. Since then, other authors have developed alternative algorithms for absorbing fixed effects, several of which have been incorporated into {\tt reghdfe}. The ideas have been extended to the Poisson model through {\tt ppmlhdfe} for Stata \citep*{correia2019ppmlhdfe}, as well as to other generalized linear models, in {\tt alpaca} and {\tt fixest} for R (\citealt{stammann2018fast, Bergé2018}). Julia packages for HDFE modeling are also available: Matthieu Gomez's FixedEffectModels.jl and Johannes Boehm's GLFixedEffectModels.jl. According to benchmarking results on the GitHub pages of {\tt fixest} and FixedEffectModels.jl, those two packages are 1--2 orders of magnitude faster than {\tt regdhfe}.\footnote{See \url{github.com/lrberge/fixest\#benchmarking} and \url{github.com/FixedEffects/FixedEffectModels.jl\#benchmarks}.}

These developments have given R and Julia users faster and more general implementations of HDFE modeling than have been available in Stata. That is what motivated the development of the {\tt julia} and {\tt reghdfejl} packages. The idea was to not reinvent wheels, but to fashion a familiar-looking Stata chassis onto existing Julia wheels.

The extended example in \cref{sec:install} followed a sequence that is the essence of {\tt reghdfejl}: copy data from Stata to Julia, estimate in Julia, return results to Stata. Here I complete the narrative arc of that example. With a small typographic change to the last Julia estimation line---wrapping a variable name in {\tt fe()}---I exploit the ability of FixedEffectModels.jl to absorb fixed effects:

\begin{stlog}
. jl: reg(auto, @formula(price ~ headroom + fe(foreign){\&}mpg), Vcov.cluster(:turn))
\smallskip
FixedEffectModel                              
============================================================================
Number of obs:                     74  Converged:                       true
dof (model):                        1  dof (residuals):                   34
R²:                             0.294  R² adjusted:                    0.264
F-statistic:                 0.991311  P-value:                        0.334
R² within:                      0.583  Iterations:                         1
============================================================================
             Estimate  Std. Error     t-stat  Pr(>|t|)  Lower 95
\HLI{76}
headroom      -273.24     274.435  -0.995646    0.3342   -855.017    308.536
(Intercept)  13743.6     2176.8     6.31369     <1e-04   9129.03   18358.2
============================================================================
\end{stlog}
Back in Stata, here is {\tt reghdfejl} running the same regression:

\begin{stlog}
. reghdfejl price headroom, absorb(foreign#c.mpg) cluster(turn)
(MWFE estimator converged in 1 iterations)
\smallskip
HDFE linear regression with Julia                 Number of obs   =         74
Absorbing 1 HDFE group                            F(   2,     15) =       0.99
Statistics cluster-robust                         Prob > F        =     0.3941
                                                  R-squared       =     0.2939
                                                  Adj R-squared   =     0.2637
Number of clusters (turn)    =         18         Within R-sq.    =     0.5829
                                                  Root MSE        =  2530.9785
\smallskip
                                  (Std. err. adjusted for 18 clusters in turn)
------------------------------------------------------------------------------
             |               Robust
       price | Coefficient  std. err.      t    P>|t|     [95\% conf. interval]
-------------+----------------------------------------------------------------
    headroom |  -273.2403   274.4352    -1.00   0.334     -855.017    308.5364
       _cons |   13743.64     2176.8     6.31   0.000     9129.026    18358.24
------------------------------------------------------------------------------
\end{stlog} 

In spirit, {\tt reghdfejl} is a slot-in replacement for {\tt reghdfe}, with the same syntax, output, and {\tt e()} return values. In practice, the two differ outwardly as well as inwardly. As {\tt reghdfjl} depends on the feature set of FixedEffectModels.jl, it offers only one optimization technique, LSMR \citep*{Fong2011-tu}---though this has never been a problem in my experience. {\tt reghdfejl} does not support ``group fixed effects." And it does not adjust the reported degrees of freedom for any collinearity of the fixed effects with each other or with the error clustering. So it sometimes reports slightly larger standard errors. While {\tt reghdfejl} can perform instrumental variables (IV) estimation, unlike {\tt ivregdfe} it does not by default work as a wrapper for {\tt ivreg2} \citep*{BSS_2007}. As a result, it can normally only perform two-stage least squares (2SLS); and for weak identification diagnosis, it only reports the Kleibergen-Paap first-stage {\it F} statistic. {\tt reghdfejl} \textit{does} accept an {\tt ivreg2} option, which directs the program to call FixedEffectModels.jl to absorb the fixed effects and then transfer control to {\tt ivreg2}. But in this mode, {\tt reghdfejl} is no faster than {\tt ivreghdfe}. For on hard problems, {\tt ivreg2} accounts for most of the run time.

{\tt reghdfe} and {\tt reghdfejl} share the following options: {\tt \underbar{a}bsorb()}, {\tt vce()}, {\tt \underbar{res}iduals()}, {\tt \underbar{tol}erance()}, {\tt \underbar{iter}ate({\textit{\#}})}, {\tt \underbar{nosamp}le}, {\tt \underbar{keepsin}gletons}, {\tt compact}, {\tt \underbar{l}evel({\textit{\#}})}, and standard Stata regression display options. In particular, {\tt reghdfejl}'s {\tt \underbar{a}bsorb()} can save some or all of the estimated fixed effects. For instance, {\tt a(firm year\_fe=year)} will absorb firm and year fixed effects, and store estimates of the latter in the variable {\tt year\_fe}. For both {\tt reghdfe} and {\tt reghdfejl}, the {\tt \underbar{tol}erance()} option sets a precision threshold for the fitting algorithms to declare convergence; however, the semantics are not necessarily the same, so that, say, {\tt tol(1e-6)} could elicit slightly different results from the two programs.

{\tt reghdfejl} accepts three distinctive options:

\hangpara {\tt gpu} requests the use of a GPU. Currently the option saves more time on NVIDIA than Apple GPUs.

\hangpara {\tt threads({\textit{\#}})} caps the number of parallel threads that can be used. It cannot increase the thread count above the limit set when Julia is started (see \cref{sec:multithreading}). This option is rarely used.

\hangpara {\tt \underbar{verb}ose} directs {\tt reghdfejl} to show more of its work---to display the Julia copy of the data set, the formula for the regression model, and the Julia regression command. The data set and formula are left behind for the user to work with\, not erased as they otherwise would be.

In addition, {\tt reghdfejl} offers a bootstrap variance-covariance estimator. {\tt vce(bootstrap)} (or {\tt vce(bs)}) is standard within Stata; {\tt regress} accepts it. As an example, for {\tt reghdfejl}, the clause

\begin{stlog}
vce(bs, reps(1000) seed(42) procs(4) cluster(year))
\end{stlog}
specifies that 1000 bootstrap replications be performed, split across four CPU threads running in parallel, with the bootstrap data drawn groupwise from groups defined by {\tt year}. One could attain the same results with {\tt bootstrap:reghdfe} or {\tt bs:reghdfe}. But {\tt reghdfejl} implements {\tt vce(bs)} in a radically faster way. It copies the primary data just once to Julia. It effects the resampling merely by updating a weight variable, so that, for example, a cluster's weights are doubled if it is drawn two times in a bootstrap sample.

To demonstrate {\tt reghdfejl}'s speed on hard problems, I create a data set with 10 million rows and two sets of fixed effects. I run the same regression with {\tt reghdfe} and {\tt reghdfejl}, as well as with the built-in {\tt areg} command, which in StataNow 18.5 gained the ability to absorb multiple fixed effects. While {\tt reghdfe} and {\tt reghdfejl} can cluster standard errors on multiple variables \citep{CGM_2011}, {\tt areg} cannot, so the example is restricted to one-way clustering:

\begin{stlog}
set obs 10000000
gen id1 = runiformint(1, 100000)
gen id2 = runiformint(1, 100)
drawnorm x1 x2
gen double y = 3 * x1 + 2 * x2 + sin(id1) + cos(id2) + runiform()
\smallskip
set processors 1
areg      y x1 x2, a(id1 id2) cluster(id1)
reghdfe   y x1 x2, a(id1 id2) cluster(id1) dof(none)
reghdfejl y x1 x2, a(id1 id2) cluster(id1)
reghdfejl y x1 x2, a(id1 id2) cluster(id1) gpu
\smallskip
set processors 6
areg      y x1 x2, a(id1 id2) cluster(id1)
reghdfe   y x1 x2, a(id1 id2) cluster(id1) dof(none)
reghdfejl y x1 x2, a(id1 id2) cluster(id1)
reghdfejl y x1 x2, a(id1 id2) cluster(id1) gpu
\end{stlog}
For comparability, the {\tt reghdfe} lines include {\tt dof(none)} to disable the degrees-of-freedom correction, since {\tt reghdfejl} lacks that feature.

The first quadruplet runs on one CPU core, as under a non-MP flavor of Stata. But that setting only affects the execution of {\tt reghdfe}'s ado and Mata code: the Julia program still runs on multiple CPU threads. On a laptop with an Intel i9-13900H CPU and an NVIDIA RTX 4070 GPU, {\tt areg} takes 8.1 seconds and {\tt reghdfe} 17 seconds. {\tt reghdfejl} takes 2.6 seconds without the {\tt gpu} option and 1.8 with. Moving to 6 processors in Stata and Mata---since the Intel chip has 6 ``performance" cores---lowers the times to 5.0, 10, 2.2, and 1.8 seconds.

Another script benchmarks {\tt reghdfejl}'s {\tt vce(bs)} option:

\begin{stlog}
webuse nlswork
xtset, clear
\smallskip
set processors 6
bs, cluster(occ\_code) reps(1000): ///
  reghdfe ln\_wage age ttl\_exp tenure not\_smsa south, absorb(year occ\_code) dof(none)
\smallskip
parallel initialize 12
parallel bs, cluster(occ\_code) reps(1000): ///
  reghdfe ln\_wage age ttl\_exp tenure not\_smsa south, absorb(year occ\_code) dof(none)
\smallskip
reghdfejl ln\_wage age ttl\_exp tenure not\_smsa south, absorb(year occ\_code) ///
  vce(bs, cluster(occ\_code) reps(1000) procs(12))
\end{stlog}
Using Stata's {\tt bs} prefix command with {\tt reghdfe} takes 69 seconds. Switching to the {\tt parallel} package of George Vega and Brian Quistorff in order to split the work among 12 copies of Stata\footnote{See \url{github.com/gvegayon/parallel}.} cuts the time to 19 seconds. {\tt reghdfejl} takes just 1.9 seconds.

The {\tt reghdfejl} package includes a {\tt partialhdfejl} command for absorbing (partialling) fixed effects out of other variables. {\tt partialhdfejl} command lines must contain either a {\tt \underbar{gen}erate()} option to name the variables that will hold the results, or a {\tt \underbar{pre}fix()} option for making the new variable names out of the old. If any of the new variables already exists, {\tt partialhdfejl} will error, unless the {\tt replace} option is also included. The following two runs produce the same point estimates:

\begin{stlog}
reghdfejl ln\_wage age ttl\_exp tenure south, absorb(year occ\_code) cluster(occ\_code)

partialhdfejl ln\_wage age ttl\_exp tenure south, absorb(year occ\_code) prefix(\_)
regress \_ln\_wage \_age \_ttl\_exp \_tenure \_south, cluster(occ\_code) nocons
\end{stlog}

To fit nonlinear HDFE models with {\tt reghdfejl}, one includes {\tt family()} and/or {\tt link()} options in the command line, following the syntax of the built-in Stata command {\tt glm}. This code below fits an HDFE Poisson model with the Mata-based {\tt ppmlhdfe} as well as with {\tt reghdfejl}. The latter runs about twice as fast:

\begin{stlog}
use http://fmwww.bc.edu/RePEc/bocode/e/EXAMPLE\_TRADE\_FTA\_DATA
egen imp = group(isoimp)
egen exp = group(isoexp)
expand 100  // expand data set 100-fold for tougher test
ppmlhdfe  trade fta, a(imp#year exp#year imp#exp) cluster(imp#exp)
reghdfejl trade fta, a(imp#year exp#year imp#exp) cluster(imp#exp) family(poisson)
\end{stlog}
In the same way, one can fit HDFE logit, binomial, negative binomial, and other models. However, at this writing, the underlying Julia package, GLFixedEffects.jl, lacks key features such as observation weighting and the equivalent of Stata's {\tt exposure()} option for Poisson models. So {\tt reghdfejl}'s ability to fit nonlinear models is currently undocumented.

\subsection{Wild bootstrapping}
\citet{RMNW_2019} introduces the Stata package {\tt boottest} for inference via the wild bootstrap. Among the methods {\tt boottest} implements is the ``wild restricted efficient" bootstrap (WRE), which is the \citet{DM_2010} adaptation of the wild bootstrap for instrumental variables (IV) estimation. Since the publication of \citet{RMNW_2019}, {\tt boottest} has been optimized in ways that especially benefit the WRE. Using the Frisch-Waugh-Lowell theorem, exogenous controls are now partialled out of the other variables using the same algebraic tricks that accelerate {\tt boottest}'s wild bootstrap for ordinary least squares. The partialling-out accelerates the application of the user's IV estimator despite its irreducible nonlinearity \citep[section 6.1]{RMNW_2019}. Also incorporated (more fully), is \citet{MACKINNON2023}'s point that the data only enter wild bootstrap computations through indexed sets of matrix products such as $\{{\bf X}'_g {\bf Y}_g : g=1,...,G\}$ where ${\bf X}_g$ and ${\bf Y}_g$ hold the rows of data matrices ${\bf X}$ and ${\bf Y}$ for cluster $g$. Reducing the data to these products early in the calculation complicates the algebra the code must implement, but speeds execution when clusters are few.

And the functionality in {\tt boottest}, so revised, has been transplanted to Julia, in the package WildBootTests.jl. Adding the {\tt julia} option to a {\tt boottest} command line switches to the new back end.\footnote{Jan Ditzen suggested this interface for calling the Julia implementation.}

Section 8.4 of \citet{RMNW_2019} approximately replicates two 2SLS regressions in the \citet{Levitt_1996} study of the short-term impact of decarceration on crime.\footnote{Data and preparation code are available at
\url{davidroodman.com/david/Levitt.zip}.} Here is a version of that example, which substitutes {\tt reghdfejl} for {\tt ivregress} and sets the stage for calling {\tt boottest}. The purpose remains to perform inference about how changes in the size of the prisoner population affect crime rates:

\begin{stlog}
use Levitt, clear
set seed 8723419
foreach crimevar in Violent Property \{
  reghdfejl D.l`crimevar{\textquotesingle}pop (LD.lpris\_totpop = ibnL.stage#i(1/3)L.substage) ///
    D.(lincomepop unemp lpolicepop metrop black a*pop) i.year, a(state) clust(state year) 
  boottest LD.lpris\_totpop, bootclust(year) ptype(equaltail) ///
    gridmin(-2) gridmax(2) nograph
\}
\end{stlog}
On a single CPU core the version of {\tt boottest} published with \citet{RMNW_2019} (version 2.3.5), needs 210 seconds to perform the two {\tt boottest} calls. The latest version finishes in 7.4 seconds, even without calling Julia. Appending {\tt julia} to the {\tt boottest} line above lowers the time to 4.1 seconds.

The R package {\tt fwildclusterboot} \citep{fwildclusterboot} also calls WildBootTests.jl to perform the WRE, and optionally does so to perform the wild bootstrap after ordinary least squares. This may be the first instance of a Julia statistical package serving as a back end on multiple platforms.

\subsection{Plotting}
\label{sec:plotting}
Many data visualization packages are available for Julia. Surveying them is beyond the scope of this article. To indicate their potential for Stata users, I generate two graphs with the plotting package Makie.jl.

\begin{figure}[t]
\includegraphics[width=\textwidth]{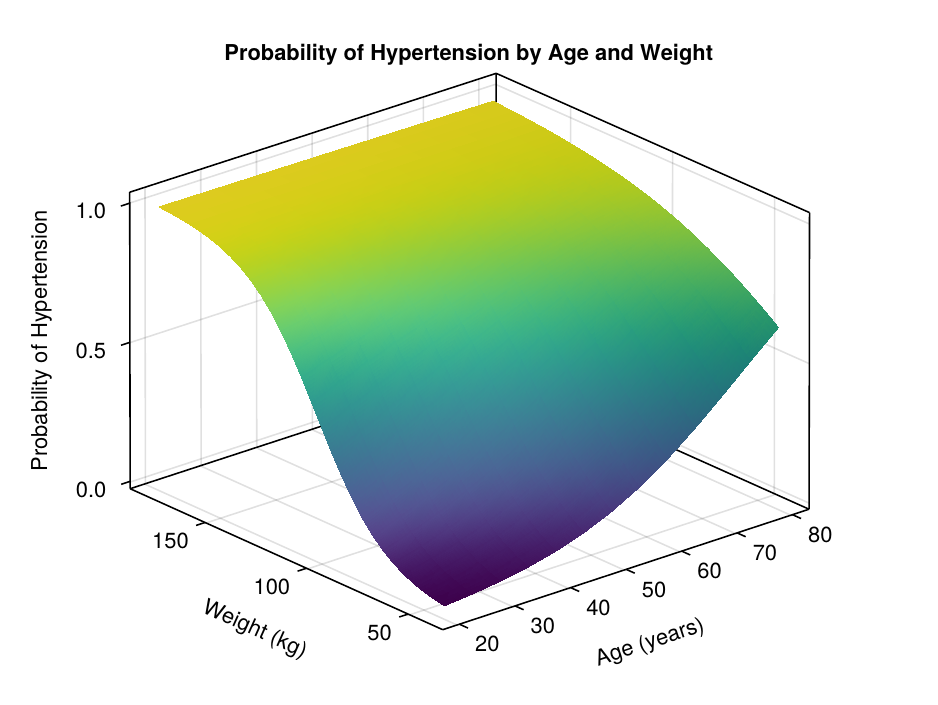}
\caption{Contour plot made with Makie in Julia}
\label{fig:EG1}
\end{figure}

Makie works with several plotting back ends, which render output in formats suited to different applications, such as interactive graphs in web pages and publication-quality images. The Cairo back end is best for the latter. It is bundled with Makie in CairoMakie.jl:

\begin{stlog}
jl AddPkg CairoMakie  // one-time installation
jl: using CairoMakie  // use now
\end{stlog}
In 2020, Chuck Huber of StataCorp blogged about using Python to visualize results from a Stata regression.\footnote{\url{blog.stata.com/2020/09/14/stata-python-integration-part-5-three-dimensional-surface-plots-of-marginal-predictions}} In Huber's example, a binary indicator of high blood pressure is regressed on age, weight, and their product using the {\tt logistic} command. {\tt margins} is then called to compute the probability of high blood pressure according to the fitted model, across a grid of ages and weights. The probabilities are copied to Python and depicted in a three-dimensional surface plot. An adaptation of that example to Julia:

\begin{stlog}
webuse nhanes2
svy: logistic highbp c.age##c.weight
margins, at(age=(20(5)80) weight=(40(5)180))
matrix xyz = r(at), r(b){\textquotesingle}  // x, y, z values in 3 columns
\smallskip
jl
  df = DataFrame(st\_matrix("xyz"), [:age, :weight, :pr\_highbp])  # get data
  f = surface(df.age, df.weight, df.pr\_highbp,
              axis=(type=Axis3,
                    title = "Probability of Hypertension by Age and Weight",
                    xlabel = "Age (years)",
                    ylabel = "Weight (kg)",
                    zlabel = "Probability of Hypertension"))
  f |> display  \# equivalent to display(f)
  save("fig1.pdf", f)
exit()
\end{stlog}
This time I format the Julia commands not as they would be written in a do file, but as they would be typed---or pasted---into Stata's Command window. That is, I drop the ``;" and ``///" from the ends of lines. The first Julia line imports the data as a matrix, converts it to a DataFrame, and names the columns. Then the surface plot is built and stored in the Julia variable {\tt f}. The call to {\tt surface()} refers to columns of the DataFrame with the convenient dot syntax, as in {\tt df.age}; here, column names are not prefixed with a colon. The next command displays the figure in a new window, which might pop up behind the Stata window. That line takes advantage of Julia's piping syntax for function calls, using the symbol ``{\tt |>}". ({\tt display(f)} is equivalent.) Last, the image is saved to disk. It appears in \Cref{fig:EG1}.

From the same data set, I plot the distribution of body mass index (BMI) by race. This time, instead of pulling the data into Julia with {\tt st\_matrix()}, I copy the entire data set from Stata with {\tt jl save}. Since I specify no name for the target DataFrame, it defaults to {\tt df}:

\begin{stlog}
jl save
jl
  dropmissing!(df, :bmi)
  races = ["Other", "White", "Black"]
  f = Figure()  \# make new, empty figure
  Axis(f[1, 1], yticks=((1:3)/20, races), xlabel="BMI")  \# add empty plot region
  for (i,r) in enumerate(races)
    density!(df.bmi[df.race .== r], offset=i/20, color=(:slategray, .4), bandwidth=1)
  end
  f |> display
exit()
\end{stlog}
{\tt dropmissing!(df, :bmi)} is similar to {\tt drop if bmi==.} in Stata. However, it also changes the type of {\tt df.bmi} from {\tt Float32?} to {\tt Float32}, meaning that the column no longer accepts missing values. That is apparently necessary to make Makie work. The ``{\tt !}'' in ``{\tt dropmissing!}" has no special meaning in the syntax of Julia: it is just a character in the function's name. But by convention, functions that modify their arguments have names ending in ``{\tt !}''.

The {\tt for...enumerate()} loop iterates over the racial groups. {\tt r} takes race names as strings while {\tt i} is a counter starting from 1. In the Stata data set, the {\tt race} column is coded with 1, 2, and 3. Those codes are displayed through a value label as ``White", ``Black", and ``Other." Since {\tt jl save} stores {\tt race} as a CategoricalVector, the Julia column's contents are outwardly strings, not numbers. {\tt df.bmi[df.race .== r]} uses Boolean indexing to extract the elements of the {\tt bmi} column for which {\tt race} takes the string value in {\tt r}.

\begin{figure}[t]
\includegraphics[width=\textwidth]{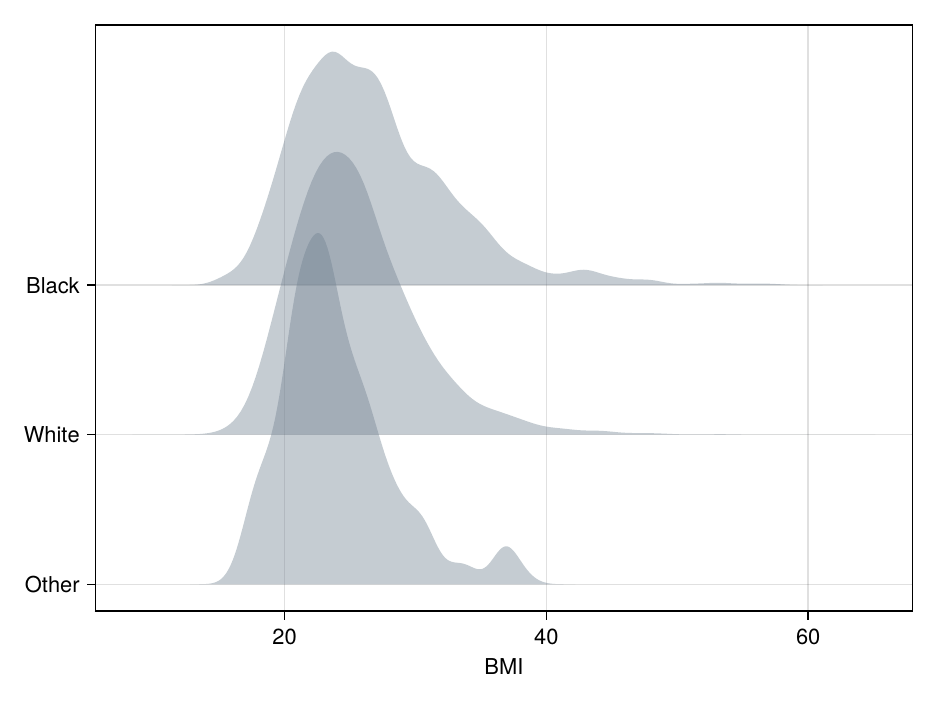}
\caption{Overlapping density plots made with Makie in Julia}
\label{fig:EG2}
\end{figure}

Each iteration of the loop plots a density in slate gray with 40\% opacity. The vertical spacing offset of $1/20$ is interpreted in the vertical axis's units, BMI density, and was chosen by trial and error to cause overlapping. See \Cref{fig:EG2}.

\section{Tips on using the {\tt jl} command}
\label{tips}

\subsection{The JuliaMono font}
Julia output routinely contains certain Unicode characters, such as vertical and diagonal ellipses, that Stata fonts do not cover. (However, this is not necessarily the case for Stata packages that call Julia back ends.) The JuliaMono font is a free, monospaced font that includes those characters, and looks reasonably good in Stata's results window. Without it, some Julia output will appear slightly garbled. The font can be installed from \url{github.com/cormullion/juliamono/releases}. Type {\tt help fonts} in Stata to learn how to change the results window font.

\subsection{Multithreading}
\label{sec:multithreading}
Julia includes several facilities for multitasking. One is distributed processing, which involves launching multiple copies of Julia that can pass messages and data to each other. A lighter form is multithreading, in which multiple execution lines run within one Julia process, largely sharing the same data space. That is the sort {\tt @tturbo} and the {\tt vce(bs)} option of {\tt reghdfejl} exploit. With regard to the latter, the maximum number of threads allowed in each Julia instance is set when the instance launches. It can default to 1. In Stata, you can determine the current limit with the command {\tt jl:Threads.nthreads()}. A good guess for the optimum is the number of ``performance" or ``big" cores on the CPU.

\textit{How} to set Julia's thread limit depends on the operating system. In Windows, you can open the Environment Variables control panel and assign {\tt JULIA\_NUM\_THREADS}. In Linux, you can add an entry such as {\tt export JULIA\_NUM\_THREADS = 6} to ``$\sim$/.bashrc" or the equivalent. In macOS, environment variables do not affect graphical applications such as Stata when launched from the desktop. One can instead set the limit in each Stata session through the {\tt threads()} option of the {\tt jl start} command. Example: {\tt jl start, threads(8)}. This must be the first {\tt jl} command issued after starting Stata. It must also precede invocation of Julia-dependent programs such as {\tt reghdfejl}.\footnote{In macOS, one can in fact assign JULIA\_NUM\_THREADS in ``$\sim$/.zshenv". This will take effect when one launches Stata from the terminal. For example {\tt open -a StataMP} will launch Stata/MP.}

\subsection{Limitations in accessing Stata objects}
\label{sec:limitations}
The {\tt julia} package lets you transfer information between Stata and Julia with commands issued in either language; \cref{syntax} will lay out more details. However initiated, the job of copying falls to the C++ plugin included in the {\tt julia} package. To read and write Stata macros, scalars, matrices, and variables, the plugin must in turn call the official Stata plugin interface.\footnote{See \url{stata.com/plugins}.} The SPI imposes certain constraints. For one, it makes public no functions that can \textit{create} Stata variables or matrices. As a result, you can edit existing Stata variables and matrices from Julia, but not create new ones. The Stata-side {\tt jl} subcommands can and do evade this constraint because they start from the Stata environment. For instance, {\tt jl GetMatFromMat} creates a destination Stata matrix before calling the plugin to copy into it from Julia.

A subtler limitation pertains to local macros. Explaining it requires a deep dive. The ``jl.ado'' file, also part of the {\tt julia} package, defines the {\tt jl} command that is invoked repeatedly in the above examples. In order to call the C++plugin, it first declares the plugin's existence with this line:

\begin{stlog}
program \_julia, plugin using(jl.plugin)
\end{stlog}
You are not supposed to call the plugin directly. But after typing that line, you could. For example:

\begin{stlog}
plugin call \_julia, eval "x=1"
\end{stlog}
The designated entry point to the plugin, the C++ function {\tt stata\_call()}, would then receive and process the two arguments after the comma, which tell it to have Julia evaluate {\tt x=1}. Since calls to the plugin are in general arcane and sometimes require preparatory steps, the {\tt jl} program serves as an essential and more user-friendly wrapper for it. That extra layer complicates access to local macros. If Julia, having been invoked by {\tt jl} and the plugin, uses the SPI function {\tt SF\_macro\_save()} to write a Stata local named {\tt foo}, then {\tt foo} will appear in the macro name space of the program that called the plugin, which is {\tt jl}. It will \textit{not} appear in the context a step up, which could be your code. As a workaround, every time the Julia function {\tt st\_local()} is called, it adds the name of the macro it is writing to another local in the name space of the {\tt jl} program, called {\tt \_\_jllocals}. Before exiting, {\tt jl} copies every macro in that list to its caller's context using the undocumented Stata command {\tt c\_local}.

However, the same complication prevents one from \textit{reading} your Stata locals from Julia. The workaround just described only goes in one direction, because I know of no way for an ado program to read its caller's local macros.

As partial compensation, single-line {\tt jl:} commands can quote locals, because Stata processes the lines before passing them to {\tt jl}. For example:

\begin{stlog}
local varname x1
jl: sum(df.`varname{\textquotesingle})
\end{stlog}
This trick does not work in interactive Julia sessions (ones started by typing {\tt jl} by itself).\footnote{The single, compound line {\tt jl: st\_local("a","123"); print("`a'")} will also fail. The second command's reference to {\tt a} will be expanded \textit{before} control is passed to Julia. The commands will work as intended if split into two lines.}

\subsection{Speed considerations in copying data: type conversions, missingness, and subsample restrictions}
Certain details of the Stata plugin interface introduce wrinkles into the copying of data between Stata and Julia.

One is that the functions for reading and writing numeric Stata variables, {\tt SF\_vdata()} and {\tt SF\_vstore()}, only truck in double-precision values, regardless of whether a Stata-side variable is stored as {\tt byte}, {\tt int}, {\tt long}, {\tt float}, or {\tt double}. By default, the {\tt jl} commands that copy from the active Stata data set to Julia DataFrames---{\tt jl save} and {\tt jl PutVarsToDF}---convert the double-precision values back to the original data types.\footnote{Because the ranges of Stata's {\tt byte}, {\tt int}, and {\tt long} types are narrowed in the way described in the next paragraph, when transferring data from Julia to Stata, the corresponding Julia types are mapped to larger types in Stata. For example, columns of type {\tt Int8}, which have an allowed range of $[-128,+127]$, are copied to Stata's two-byte {\tt int} type, since Stata {\tt byte}s only hold $[-128,100]$.} So a Stata {\tt byte} variable will be converted to {\tt double} and back on the way to appearing in a Julia DataFrame. The {\tt jl} commands can do nothing about the first conversion since that is performed by Stata. But they do accept a {\tt \underbar{double}only} option to skip the second conversion and instead leave all copied values in double precision. This typically saves time, by reducing conversions and allowing Julia to allocate a contiguous block of memory to receive purely double-precision values, rather than a separate block for each of variously typed columns.

Separately, Stata and Julia represent missing values in different ways. Stata reserves certain high values of each data type for 27 flavors of missing: {\tt .} and {\tt .a}, ..., {\tt .z}. That is why the maximum value of a Stata {\tt byte} is 100, not than 127. Julia only supports one flavor of missing, and stores missingness information alongside rather than in data values. It thus distinguishes between empty and missing strings. The {\tt jl} subcommands translate missings when copying between Stata data sets and Julia DataFrames (but not Julia matrices). For data known to contain no missing values, adding the {\tt \underbar{nomiss}ing} option to the copying commands will save time by skipping this step. 

Finally, the SPI lets plugin calls include {\tt \textit{if}} or {\tt \textit{in}} clauses to restrict the Stata sample on which the commands will operate. When such clauses are \textit{not} specified, the {\tt julia} plugin runs specially optimized code.

Data transfers are thus fastest when they map double precision to double precision and include no missing values or sample restrictions. Consider this example, which creates a 100 million $\times$ 10 set of double-precision numbers and copies them to a Julia DataFrame named {\tt demo}:

\begin{stlog}
set obs 100000000
drawnorm double x1-x10
jl PutVarsToDF x1-x10, nomissing doubleonly dest(demo)
\end{stlog}
On my computer, the last line takes 1.3 seconds, for a transfer rate of 6 gigabytes/second.

\subsection{Simulated read-eval-print loop}
The Julia language system has many components---for compiling code, managing memory, and so on. The core parts are written in C++ and distributed as compiled libraries. Other components are themselves written in Julia, including the front end, which users often think of as Julia per se. The front end is called the REPL, for ``Read-Eval-Print Loop." As its name makes plain, the REPL's main job is to read in commands, evaluate them, print the results, and repeat. But the REPL provides other conveniences. Typing "?" starts help mode. Typing "]" instead switches to the Julia package manager. Greek letters and other symbols can be entered through a TeX-like syntax. Arrow keys allow one to explore one's command history.

While there may be a way to run the full Julia REPL inside of Stata, the {\tt jl} command does not do that. It gins up a limited REPL. That is because the plugin directly accesses lower-level components of Julia, through the C-language entry points that loosely comprise Julia's interface for ``embedding" in C. A workhorse for {\tt jl} is the Julia system's {\tt jl\_eval\_string()} function. It accepts a C string holding a Julia expression, evaluates it, and returns the result in a C data structure that represents a generic Julia object. All Julia lines submitted to {\tt jl} pass through this function. As its name suggests, the function's job is to evaluate strings, not provide a user experience.

Especially when in interactive mode, {\tt jl} takes steps to create an ersatz Julia REPL. As described in the first example in \cref{firstsession}, it checks whether lines are syntactically complete and prompts the user to continue typing if not. That allows one to enter a multiline code block without triggering a syntax error when it is still incomplete. {\tt jl} also captures printed output (from {\tt stdout}) in order to display not only the return value of a command but anything printed while running it. {\tt jl} translates lines beginning with ``?" into help requests and suppresses output from lines ending with ``;".

Another REPL feature that {\tt jl} simulates matters for code such as this:

\begin{stlog}
S = 0
for i in 1:10
  S += i
end
print(S)
\end{stlog}
This snippet prints the sum of the first 10 whole numbers---at least when run in the Julia REPL. Actually, for reasons having nothing to do with Stata, the snippet will crash if run non-interactively, as part of a stored Julia program file. Under Julia's ``soft scoping" rule, since {\tt S} is declared as a global variable outside the for loop or any other structure such as a function definition, it is invisible within the for loop. The first attempt to increment it from within the loop will trigger an undefined-variable error. Because that rule is counterintuitive, Julia's REPL engineers an exception for interactive use.\footnote{See the documentation at \url{docs.julialang.org/en/v1/manual/variables-and-scoping/\#local-scope}.} {\tt jl} imitates the exception by calling the Julia {\tt softscope()} function on the user's input after parsing it. {\tt softscope()} is itself part of the code base of the Julia REPL.\footnote{Thanks to Mark Kittisopikul for the idea of calling {\tt softscope()}.}

Still, {\tt jl}'s REPL simulation is incomplete. Typing {\tt ?} and {\tt ]} alone have no effect. Plots do not automatically appear, which is why the examples in \cref{sec:plotting} explicitly call {\tt display()}.

And the processing required for the simulation, along with the special handling of Stata locals, adds time to the execution of each Julia line from Stata. The extra 0.02 seconds or so is not noticeable in an interactive session. But it can add up in a program that calls {\tt jl:} many times. The {\tt \_jl:} variant of the prefix command saves time by bypassing all of this REPL simulation.

\section{The {\tt julia} package}
\label{syntax}

\subsection{Architecture}
The {\tt julia} package consists of three intertwined pieces. The core is a shared library written in C++ according to the specification for Stata plugins. It comes precompiled for Windows, Linux, and macOS with Intel or Apple silicon. At installation, the ``julia.pkg" manifest file instructs Stata to select the copy of the library that will run on the target computer, and to name it ``jl.plugin." The plugin performs two main jobs: efficiently copying data among matrices and data sets in Julia and Stata, and sending user-written commands to Julia and returning the results.

The file ``jl.ado'' serves as a Stata front end for the plugin, providing a more intuitive interface and performing tasks better done in the ado language than in C++. At the suggestion of Jeffrey Pitblado at StataCorp, this program is called {\tt jl} rather than {\tt julia} because Stata might introduce an official {\tt julia} command. When {\tt jl} is first called in a Stata session, it briefly runs the true Julia REPL through a shell command in order to query the location on disk of Julia's shared libraries. It passes the location to the plugin, which loads the libraries.\footnote{Thanks to David Anthoff and Elliot Saba for guidance on this solution.}

{\tt jl} also calls Stata's {\tt findfile} to track down the third key piece of the package. ``stataplugininterface.jl" defines a module of Stata interface functions in Julia, such as {\tt st\_numscalar()} and {\tt st\_matrix()}, which are used in examples above.\footnote{It would be more ``Julianic" to name the file ``StataPluginInterface.jl". But rather than being posted to Julia's general package registry, the file is distributed as part of a Stata package, and the Stata package installer cannot handle uppercase characters.} These work by calling C-language entry points in the Stata plugin interface, which are exported to Julia by the plugin. (The Julia language includes a way to call C functions.) At start-up, {\tt jl} therefore stores the location of ``jl.plugin" in a variable in the loaded stataplugininterface module.

\subsection{Stata commands in the {\tt julia} package}
As we have seen, {\tt jl:} and the faster {\tt \_jl:} prefix commands run lines of Julia code:

\begin{stlog}
jl: \textit{juliaexpr}
\_jl: \textit{juliaexpr}
\end{stlog}
where \textit{juliaexpr} is an expression to be evaluated in Julia.

{\tt jl} also accepts subcommands. As discussed in \cref{sec:multithreading}, one is especially useful in macOS, because it allows users to set the number of available threads:

\begin{stlog}
jl start [, threads(# | auto)]
\end{stlog}
Specifying {\tt threads(auto)} instructs Julia to use its own heuristic to pick the number of threads. Executing {\tt jl start} is optional, for {\tt jl} will initialize itself as needed.

Several subcommands help manage Julia packages, which are discrete libraries of code structured for easy sharing and are usually hosted on GitHub. Just as Stata downloads packages into folders such as ``/home/droodman/ado/plus," Julia downloads and stores packages in directories called \textit{environments}. Julia also lets you create new environments and switch among them. This is useful for the developer of a Julia-calling Stata program because it allows the Stata program to install needed Julia packages without interfering with the user's package configuration. The package management subcommands are:

\begin{stlog}
jl GetEnv
jl SetEnv [\textit{name}]
jl AddPkg \textit{name}, [minver(\textit{X.Y.Z})]
\end{stlog}
{\tt GetEnv} displays the directory and contents of the current package environment and returns them in {\tt r()} macros. {\tt jl SetEnv} moves to a chosen, named environment, which is merely a subdirectory of the default environment directory. If the environment does not exist, it is created and populated with the DataFrames.jl and CategoricalArrays.jl packages. One can return to the default environment with ``{\tt SetEnv .}" or just ``{\tt SetEnv}".\footnote{In distributed computing, Julia worker processes do not automatically inherit the master's package environment. One can pass on the master's environment through command line switches when launching the workers, such as in {\tt addprocs(4, exeflags="--project=\$(Base.active\_project())")}.}

On my computer, creating a new environment called {\tt myenv} produces this output:

\begin{stlog}
. jl SetEnv myenv
Current environment: myenv, at /Users/davidroodman/.julia/environments/v1.11/myenv

Status '~/.julia/environments/v1.11/myenv/Project.toml'
[324d7699] CategoricalArrays v0.10.8
[a93c6f00] DataFrames v1.7.0
\end{stlog}

Last in the package management family is the {\tt AddPkg} subcommand. Within the current environment, the subcommand installs a package; or, if {\tt minver()} is specified, updates the package to the latest version if the current version falls below the specified minimum. Versions \textit{X.Y.Z} are in the three-part semantic versioning format.

The remaining {\tt jl} subcommands copy data between Stata and Julia. The first six listed below are meant more as programmer's commands while the last two are higher-level.\footnote{Johannes Boehm suggested adding high-level data-copying commands.} {\tt jl GetVarsFromDF}, for example, requires that the Stata data set already be at least as tall as the DataFrame from which it will receive variables. In contrast, {\tt jl use} works like Stata's {\tt use} command: it replaces the current data set, and will fail unless the {\tt replace} option is included, or the current data set has not changed since it was last saved.

\begin{stlog}
jl PutVarsToDF [\textit{varlist}] [\textit{if}] [\textit{in}], [\underbar{dest}ination(\textit{string}) cols(\textit{string}) \underbar{nolab}el
                                     \underbar{nomiss}ing \underbar{double}only]
jl GetVarsFromDF \textit{varlist} [\textit{if}] [\textit{in}], [cols(\textit{string}) source(\textit{string}) replace \underbar{nomiss}ing]
jl PutVarsToMat [\textit{varlist}] [\textit{if}] [\textit{in}], \underbar{dest}ination(\textit{string})
jl GetVarsFromMat \textit{varlist} [\textit{if}] [\textit{in}], source(\textit{string}) [replace]
jl PutMatToMat \textit{matname}, [\underbar{dest}ination(string)]
jl GetMatFromMat \textit{matname}, [source(\textit{string})]
\smallskip
jl save [\textit{dataframename}], [\underbar{nolab}el \underbar{nomiss}ing \underbar{double}only]
jl use \textit{dataframename}, [clear]
jl use \textit{varlist} using \textit{dataframename}, [clear]
\end{stlog}

The \textit{varlist}s and \textit{matname}s before the commas refer to Stata variables or matrices. If an optional \textit{varlist}
is omitted, it defaults to {\tt *}, meaning all variables in the current data set in their current order. A {\tt \underbar{dest}ination()} or {\tt source()} option after the comma should contain the name of a Julia matrix or DataFrame. When an optional DataFrame name is not provided, it defaults to {\tt df}. The {\tt cols()} option specifies the DataFrame columns to be copied to or from. It defaults to the Stata \textit{varlist} before the comma. Destination Stata matrices and Julia matrices and DataFrames are entirely replaced. The low-level commands will create the destination Stata variables or, if {\tt replace} is specified, overwrite them subject to any {\tt [\textit{if}]} or {\tt [\textit{in}]} restriction.

\subsection{Julia functions in the {\tt julia} package}
Functions in the stataplugininterface.jl module can also be characterized as low- or high-level. The low-level ones wrap and mimic functions in the Stata plugin interface. For example, {\tt SF\_vdata()} extracts a single observation on a single Stata variable. The higher-level ones build on the low-level ones in order to imitate some of Mata's Stata interface functions. {\tt st\_view()}, for instance, returns a Julia view onto one or more numeric Stata variables, optionally restricting to a subsample. Through the view, Julia code can read and write Stata variables while treating them as part of a matrix.

The full list of Stata-interfacing Julia functions follows:

\begin{tabular}{ll}
      {\tt SF\_nobs()} & Number of observations in Stata data set \\
      {\tt SF\_nvar()} & Number of variables in Stata data set \\
      {\tt SF\_var\_is\_string(i)} & Whether variable {\tt i} is string \\
      {\tt SF\_var\_is\_strl(i)} & Whether variable {\tt i} is a strL \\
      {\tt SF\_var\_is\_binary(j, i)} & Whether observation {\tt j} of variable {\tt i} is a binary strL \\
      {\tt SF\_sdatalen(j, i)} & String length of variable {\tt i}, observation {\tt j} \\
      {\tt SF\_is\_missing()} & Whether a {\tt Float64} value is Stata missing \\
      {\tt SF\_missval()} & Stata floating-point value for missing \\
      {\tt SF\_vstore(j, i, val)} & Set observation {\tt j} of variable {\tt i} to {\tt val} (numeric) \\
      {\tt SF\_sstore(j, i, s)} & Set observation {\tt j} of variable {\tt i} to {\tt s} (string) \\
      {\tt SF\_vdata(j, i)} & Get observation {\tt j} of variable {\tt i} (numeric) \\
      {\tt SF\_sdata(j, i)} & Get observation {\tt j} of variable {\tt i} (string) \\
      {\tt SF\_macro\_save(mac, tosave)} & Set macro {\tt mac} \\
	  {\tt SF\_macro\_use(mac)} & Contents of macro {\tt mac} \\
	  {\tt SF\_scal\_save(scal, val)} & Set value of scalar {\tt scal}
    \end{tabular}
      
\begin{tabular}{ll}
		{\tt SF\_scal\_use(scal)} & Get scalar {\tt scal} \\
		{\tt SF\_row(mat)} & Number of rows of matrix {\tt mat} \\
		{\tt SF\_col(mat)} & Number of columns of matrix {\tt mat} \\
		{\tt SF\_mat\_store(mat, i, j, val)} & {\tt mat[i,j] = val} \\
		{\tt SF\_mat\_el(mat, i, j)} & Get {\tt mat[i,j]} \\
		{\tt SF\_display(s)} & Print to Stata results window \\
		{\tt SF\_error(s)} & Print error to Stata results window \\
		{\tt st\_nobs()} & Same as {\tt SF\_nobs()} \\
		{\tt st\_nvar()} & Same as {\tt SF\_nvar()} \\
      {\tt st\_varindex(s)} & Index of variable named {\tt s} in data set \\
      {\tt st\_global(mac)} & Get global macro {\tt mac} \\
      {\tt st\_global(mac, tosave)} & Set global macro {\tt mac} \\
      {\tt st\_local(mac, tosave)} & Set local macro {\tt mac} \\
      {\tt st\_numscalar(scal)} & Get scalar {\tt scal}; same as {\tt SF\_scal\_use()} \\
      {\tt st\_numscalar(scal, val)} & Set scalar scal; same as {\tt SF\_scal\_save()} \\
      {\tt st\_matrix(matname)} & Get numeric Stata matrix \\
      {\tt st\_matrix(matname, jlmat)} & Put Julia matrix in existing Stata matrix \\
      {\tt st\_data(varnames)} & Get Stata variables, as matrix \\
      {\tt st\_data(varnames, sample)} & Get Stata variables, with sample restriction, as matrix \\
      {\tt st\_view(varnames)} & Get Stata variables, as view \\
      {\tt st\_view(varnames, sample)} & Get Stata variables, with sample restriction, as view 
\end{tabular}

The limitations noted in \cref{sec:limitations} can be seen in the list. While {\tt st\_global()} can read and write globals, {\tt st\_local()} can only write locals. {\tt st\_matrix()} can only write to Stata matrices that already exist.

An interactive session demonstrates the use of some of the functions:

\begin{stlog}
sysuse auto
jl
  sample = rand(Bool, st\_nobs())    # random Boolean vector defining subsample
  v = st\_view("price mpg", sample)  # view onto subsample of two Stata variables
  v ./= 2                           # halve these values in the Stata data set
  st\_numscalar("s", sum(v))         # sum the halved data into Stata scalar s
  st\_local("m", string(st\_numscalar("s"))) # put sum as string in Stata local m
exit()
\end{stlog}
One can obtain somewhat more information about each function through the help facility, for example with ``{\tt jl:?st\_view}".

\section{Conclusions}
\label{conclusion}
Julia is a strong option for universal back-end development of statistical applications. True, working in Julia comes with drawbacks, including the challenge of learning a complex language, the delays upon installation and first use of programs, and the immaturity of some packages and their documentation. But Julia and nearly all Julia packages are free and open and run on all popular platforms. The language blends features for maximizing performance where it matters with a Python-like expressiveness that makes code easier to write and maintain. The package ecosystem covers domains essential to efficient statistical work, such as multiprocessing, linear algebra, and numerical optimization. 

Instead of implementing new methods separately in Stata or R or Python, or in a low-level language such as C, developers can write the back ends once, in Julia. Here, WildBootTests.jl shows the way: it is an optional back end for {\tt boottest} in Stata and {\tt fwildclusterboot} in R, and a required back end for instrumental variables inference with the latter.

For Stata users and developers, the {\tt julia} package demonstrates the practicality and value of bridging from Stata to Julia. Through a plugin, data can be piped at high speed. Most features needed to make Stata objects accessible in Julia can and have been implemented. The linkage gives Stata users access to Julia packages for estimation and plotting that offer greater speed or new features. {\tt reghdfejl} might be the ``killer app'' for the Julia-Stata link. It fits linear models much faster than the pioneering {\tt reghdfe}, at least for problems that are time consuming for the latter. It also outpaces the newly augmented {\tt areg}. And, preliminarily, {\tt reghdfejl} handles nonlinear HDFE models, thus bringing new functionality to Stata users.

StataCorp could tighten Stata's integration with Julia, as it has done with Python and Java. Stata's developers could, for example, extend the plugin interface to allow creation of Stata matrices and variables. They could add a Julia mode to the do file processor, so that Julia code could be embedded without awkward semicolons and line continuation symbols. They might be able to grant more access to Stata variable storage \textit{en bloc}, so that Julia code could read and write Stata data without copying it one observation at a time, or even without copying it at all. They could give Mata the ability to call Julia functions.

The goal for the {\tt julia} package was to prove a concept. In that I think it has succeeded.

\section{Acknowledgments}

I thank Sergio Correia, Páll Haraldsson, and an anonymous reviewer for comments.

\bibliographystyle{sj}
\bibliography{julia}

\begin{aboutauthor}
	David Roodman is a senior advisor at Open Philanthropy. He wrote {\tt xtabond2}, {\tt cmp}, {\tt boottest}, and other Stata programs. He received the first \textit{Stata Journal} Editors' Prize, in 2012.
\end{aboutauthor}

\clearpage
\end{document}